\documentclass[%
 reprint,
 amsmath,amssymb,
 aps,
]{revtex4-2}
\usepackage{graphicx}
\usepackage{dcolumn}
\usepackage{bm}

\usepackage{xcolor}
\begin{document}
\preprint{APS/123-QED}

\title{Topological Dirac Spin-Gapless Materials - New Horizon for Topological Spintronics Without Spin-Orbit Interaction}

\author{Muhammad Nadeem}
\email{mnadeem@uow.edu.au}
\affiliation{Institute for Superconducting and Electronic Materials (ISEM), Australian Institute for Innovative Materials (AIIM), University of Wollongong, Wollongong, New South Wales 2525, Australia.}
\affiliation{ARC Centre of Excellence in Future Low-Energy Electronics Technologies (FLEET), University of Wollongong, Wollongong, New South Wales 2525, Australia}
\author{Xiaolin Wang}
\email{xiaolin@uow.edu.au}
\affiliation{Institute for Superconducting and Electronic Materials (ISEM), Australian Institute for Innovative Materials (AIIM), University of Wollongong, Wollongong, New South Wales 2525, Australia.}
\affiliation{ARC Centre of Excellence in Future Low-Energy Electronics Technologies (FLEET), University of Wollongong, Wollongong, New South Wales 2525, Australia}

\begin{abstract}
The existence of chiral edge states, corresponding to the nontrivial bulk-band topology characterized by a non-vanishing topological invariant, and the manipulation of topological transport via chiral edge states promise topological electronic/spintronic device applications. Here we predict the existence, practical realization, topological protection, and topological switching of spin-gapless valley-filtered chiral edge states, representing a novel topological Dirac spin-gapless/half-metal phase in antiferromagnetic honeycomb structures terminated on zigzag edges. We demonstrate that this phenomenon is realizable if a perpendicular (transverse) electric field is applied in zigzag nanoribbons with an antiferromagnetic ordering on the boundary (in the bulk), and the Weber-Fechner type nonlinear behavior is optimizable by a transverse (perpendicular) electric field. The existence of spin-gapless valley-filtered chiral edge states, their correspondence with nontrivial topological character in the bulk, and electric-field-driven switching of their spin-polarization that is accompanied by switching of bulk-band topology promise a new strategy for topological spintronics without spin-orbit interaction.
\end{abstract}
\maketitle


\section{\label{intro}Introduction}
The existence of topologically protected edge states along one-dimensional (1D) boundary, owning to the bulk band topology of a two-dimensional (2D) system, has brought about an interest in the Berry-phase supported dissipationless topological transport \cite{RMPBerry,TBT,Song17}. The nature of edge states in 2D topological insulating materials is characterized through a specific non-vanishing topological invariant in the bulk. For instance, integer quantum Hall (IQH) \cite{IQH} and quantum anomalous Hall (QAH) \cite{Haldane,Liu08T1,Wang14T2,Yu10,Chang13,Qiao10,Tse11,Qiao12,Qiao14-G/AF,Babar19,Nadeem20} , quantum spin Hall (QSH) \cite{Kane05a,Kane05b,QSH-Bernevig,Bernevig06,Konig07,Liu-typeII,Knez-TypeII,Du-TypeII,QV-Qiao-BLG,Qian14,Wu,TMDC-1H}, quantum valley Hall (QVH) \cite{QV-Xiao-G,QV-Yao,QV-Qian18,BLG-Min,Igor20}, and quantum spin-valley Hall (QSVH) \cite{Xiao-SV-AFM,ISLAM16,Igor21} with coupled spin and valley degrees of freedom \cite{SV-Xiao-TMDC,SV-Zhou-TMDC} are well known examples of 2D topological insulators. These novel systems are distinct from conventional insulators; though the current cannot flow via bulk states as the empty bands are separated from fully occupied bands by an energy gap, however, a quantized conduction may still be allowed via edge states lying inside the energy gap. Such a nontrivial energy gap is characterized by Chern number ($\mathcal{C}$) in IQH/QAH insulators \cite{TKNN,Haldane,Tse11} with chiral edge states (CES), spin Chern number ($\mathcal{C}_s$) in QSH insulators \cite{Kane05a,Haldane06,3DTI-FuZ2} with spin-filtered (helical) chiral edge states (SF-CES), valley Chern number ($\mathcal{C}_v$) in QVH insulators \cite{Tse11,QV-Qiao-BLG,Fan13,QV-Qian18} with valley-filtered chiral edge states (VF-CES), and spin-valley Chern number ($\mathcal{C}_{sv}$) in QSVH insulators \cite{Xiao-SV-AFM,Igor21} with spin-valley-filtered chiral edge states (SVF-CES). Such nontrivial bulk and dissipationless chiral edge states may provide promising platform for topological device applications \cite{Topo-Elec,Topo-spin,Topo-Val07,Topo-Val16,phdthesis}.\par

\begin{figure}
\includegraphics[scale=0.47]{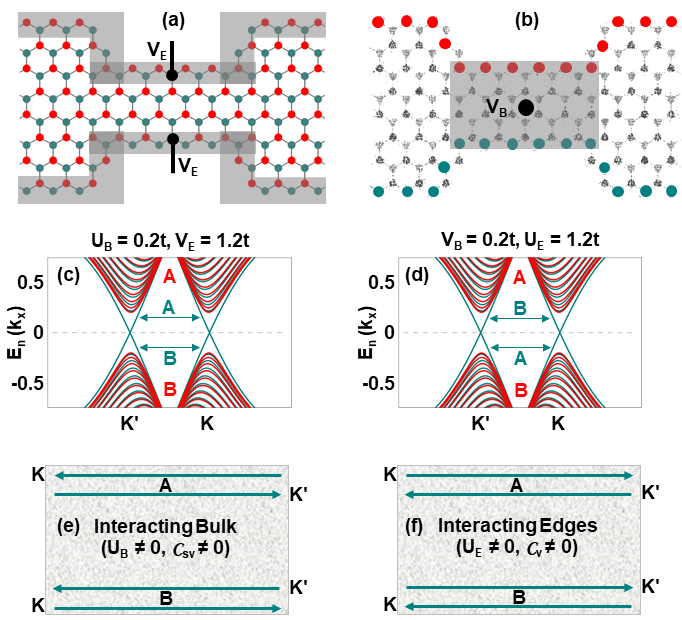}
\caption{\label{BBC}\textbf{Topological Dirac spin-gapless phase hosting SG-VF-CES in ZNRs of 2D antiferromagnets}. \textbf{(a,b)} ZNRs with antiferromagnetic ordering in the bulk (a) and on the boundary (b) connected with side gates and top/bottom gates respectively. \textbf{(c,d)} 1D band dispersion for 32-ZNRs showing SG-VF-CES for geometries depicted in (a) and (b). \textbf{(e,f)} Schematic representation of SG-VF-CES in real space. Here gray and red/cyan colours represent non-magnetic and magnetic atoms with spin up/down moments respectively.}
\end{figure}

Unlike QSH and QAH insulators where strong spin-orbit interaction (SOI) is desired to induce CES, VF-CES and SVF-CES can be realized through different physical mechanisms in QVH and QSVH insulators, leading to unconventional bulk-boundary correspondence \cite{Li10}. For instance, by engineering topological domain walls \cite{Semenoff08,QV-Martin-BLG, QV-Qian18,Xiao-SV-AFM,Igor21}, inducing ferromagnetic exchange interaction \cite{Tse11}, and by applying perpendicular electric field (PEF) \cite{Castro08} or transverse electric field (TEF) \cite{QV-Wang-G,QV-Qian18}. While electrical domain walls \cite{Semenoff08,QV-Martin-BLG,QV-Qian18} and/or TEF \cite{QV-Wang-G,QV-Qian18} induce spin-degenerate VF-CES in nonmagnetic QVH insulators, magnetic domain walls in QSVH insulators \cite{Xiao-SV-AFM} and an interface between QVH and QSH insulators \cite{Igor21} can induce SVF-CES. Unlike SF-CES in QSH insulators that are protected by time-reversal symmetry, topological robustness of QVH and QSVH insulators is guaranteed by large inter-valley separation in momentum space, and thus, suppressed inter-valley scattering \cite{Morozov06, Morpurgo06,Gorbachev07}. \par

\section{\label{Result}Results and discussion}
Here we discover a new strategy that exploits both TEF and PEF and promise the existence, practical realization, topological protection, and topological switching of spin-gapless valley-filtered chiral edge states (SG-VF-CES) in zigzag nanoribbons (ZNRs) of 2D antiferromagnets. SG-VF-CES are VF/SVF-CES displaying spin-gapless Dirac dispersion \cite{SGS1,SGS2} where all the edge states lying inside the nontrivial energy gap carry same spin-polarization, representing a novel topological Dirac spin-gapless/half-metal phase. By calculating tight binding band dispersion, density of states (DOS), and edge state wave functions for ZNRs with an antiferromagnetic ordering on the boundary (in the bulk), it is demonstrated that this phenomenon is realizable through a PEF (TEF) and the Weber-Fechner type nonlinear behavior \cite{Bhowmick10} is optimizable through a TEF (PEF).\par

\begin{figure}
\includegraphics[scale=0.35]{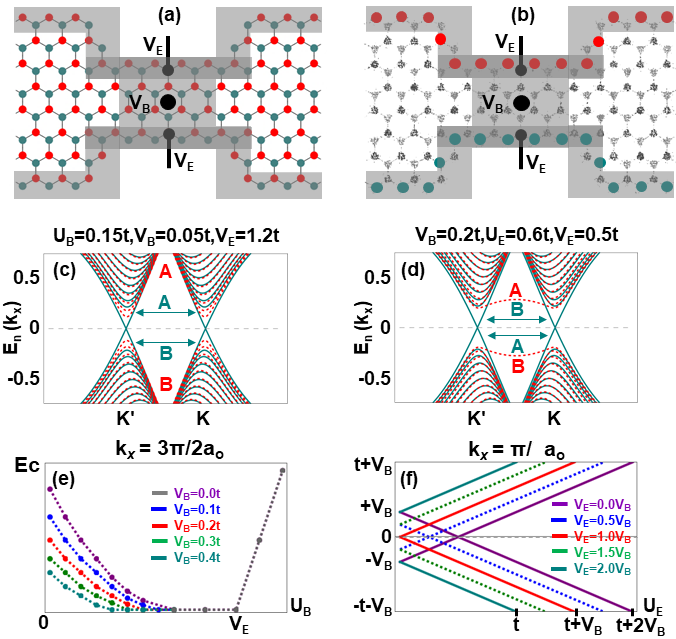}
\caption{\label{BBC-1}\textbf{Topological Dirac spin-gapless phase with bulk-boundary gating scheme.} \textbf{(a)} In ZNRs with bulk magnetism, boundary gates induce SG-VF-CES while bulk gates optimize the critical strength of bulk magnetism. \textbf{(b)} In ZNRs with boundary magnetism, bulk gates induce SG-VF-CES while boundary gates optimize the critical strength of boundary magnetism. \textbf{(c,d)} 1D band dispersion for 32-ZNRs showing SG-VF-CES for geometries depicted in (a) and (b). \textbf{(e,f)} Energy dependence of confined boundary states, associated with momentum $k_x=3\pi/2$ and $k_x=\pi$ respectively, on the bulk (e) and the boundary (f) magnetism. The critical strength of bulk magnetism required for closing the confinement induced energy gap at valleys (e) and the critical strength of boundary magnetism required for complete submergence of edge states with the bulk subbands at $k_x=\pi$ (f) is decreased by the application of perpendicular and transverse electric field respectively. Note that, in (e), the energy gap shifts to the momentum $k_x=\pi$ when $U_B>V_E$, depicted by gray colour. In (e), tight binding calculations are performed for 6-ZNRs with $V_E=1.2t$ while the results presented in (f) are independent of the ribbon's width.}
\end{figure}

Similar to the Weber-Fechner type nonlinear response that depends logarithmically on the boundary potentials \cite{Bhowmick10}, we note that the dependence of energy gaps on staggered potentials (SPs) in the bulk of quantum confined ZNRs is also nonlinear. Moreover, the inter-edge antiferromagnetic configuration is the lowest-energy ground state only for small-size systems (i.e., $N<32$) \cite{HMGNRs}, the antiferromagnetically ordered phase exist only when the strength of SSPs in the bulk is smaller than a critical limit \cite{Soriano12-G}, and unrealistically large (small) strength of SPs in the bulk (on the boundary) may lead to non-quantized Hall conductance due to mixing of inter-valley Berry curvature \cite{Lee19} (intra-valley modes \cite{Qiao11-G}). It implies that a state of the art optimization scheme is desired for the practical realization and topological robustness of SG-VF-CES. In light of this, unlike previously proposed schemes that utilize either TEF or PEF for inducing edge state half-metallicity in graphene ZNRs \cite{HMGNRs,Soriano12-G,Qiao11-G, LEE17-G,LEE18-G}, proposed strategy that exploits both TEF and PEF promises practical realization of SG-VF-CES, guarantees large Fermi velocity, high mobility, and large signal-to-noise ratio owning to the Dirac dispersion, and ensures topological robustness against both long-range and short-range disorders \cite{SI}.\par

In both QVH and QSVH phases, with antiferromagnetic ordering on the boundaries (with unit cell length scale) and in the bulk respectively, spin-transport via SG-VF-CES can be switched by a PEF. Interestingly, PEF-driven switching of spin-transport is accompanied by the switching of bulk topological invariant, a blueprint for topological switching. For instance, in QVH insulators, reversal of PEF switches spin-polarization of SG-VF-CES by switching the valley Chern number from $\mathcal{C}_v=\pm2$ to $\mathcal{C}_v=\mp2$. Similarly, in QSVH insulators, PEF switches SG-VF-CES to VF-CES by switching bulk-band topology characterized by $\mathcal{C}_{sv}=-2$ to bulk-band topology characterized by $\mathcal{C}_v=-2$. The existence of SG-VF-CES in nanometre-scale ZNRs, the topological origin of their spin-polarization, and electric field-driven topological switching of their spin-polarization promise a new strategy for topological spintronics, even in the absence of SOI. First, the existence of SG-VF-CES allows utilizing topological Dirac spin-gapless materials both as a spin injector/detector and spin-transport channel with large spin-relaxation lifetimes and spin-diffusion distances. Second, topological switching of spin-polarization makes our proposals for topological spin field effect transistors (TSFETs) contrasting from the existing models in which SOI is a key ingredient \cite{TSFET-Krueckl,TSFET-Zhang, TSFET-Dolcini,TSFET-Battilomo,TSFET-Acosta}.\par

In N-ZNRs composed of N zigzag chains, with antiferromagnetic ordering in the bulk or on the boundary and PEF ($E_z$) and/or TEF ($E_y$), SG-VF-CES traversing along x-axis can be characterized by a nearest-neighbour tight-binding Hamiltonian
\begin{widetext}
\begin{equation}
H_{U_E+E_y}^{E_z} = t\sum_{\langle ij\rangle\alpha}c_{i\alpha}^\dagger c_{j\alpha}+\sum_{i\alpha} c_{i\alpha}^\dagger \left(v_iV_B\right) c_{i\alpha}+\sum_{i(=1\land N)\alpha} c_{i\alpha}^\dagger \left(s_z\tilde{u}_iU_E+\tilde{v}_iV_E\right) c_{i\alpha}\;.\label{UE}
\end{equation}
\begin{equation}
H_{U_B+E_z}^{E_y}= t\sum_{\langle ij\rangle\alpha}c_{i\alpha}^\dagger c_{j\alpha}+\sum_{i\alpha} c_{i\alpha}^\dagger \left(s_zu_iU_B+v_iV_B\right) c_{i\alpha}+\sum_{i(=1\land N)\alpha} c_{i\alpha}^\dagger \left(\tilde{v}_iV_E\right) c_{i\alpha}\;.\label{UB}
\end{equation}
\end{widetext}
where $c_{i\alpha}^\dagger (c_{i\alpha})$ is the creation (annihilation) electron operator with spin polarization $\alpha=\uparrow,\downarrow$ on site \textit{i} and $s_z$ represents the electron intrinsic spin. First term represents nearest-neighbour hopping, while second and third terms incorporate on-site energies due to staggered sublattice potentials (SSPs) and staggered magnetic potentials (SMPs). In the bulk, $V_B$ ($U_B$) represent SSPs (SMPs) with $v_i=u_i=+1(-1)$ for sublattice A(B). On the boundary, $V_E$ ($U_E$) represent SSPs (SMPs) with $\tilde{v}_{1(N)}=\tilde{u}_{1(N)}=+1(-1)$ for left(right) boundary terminated on sublattice A(B) while $\tilde{v}_i=\tilde{u}_i=0$ for all other bulk sites. Electric fields in superscript are required for the existence of SG-VF-CES while those in the subscript are desired for the optimization of SMPs.\par

\begin{figure}
\includegraphics[scale=0.415]{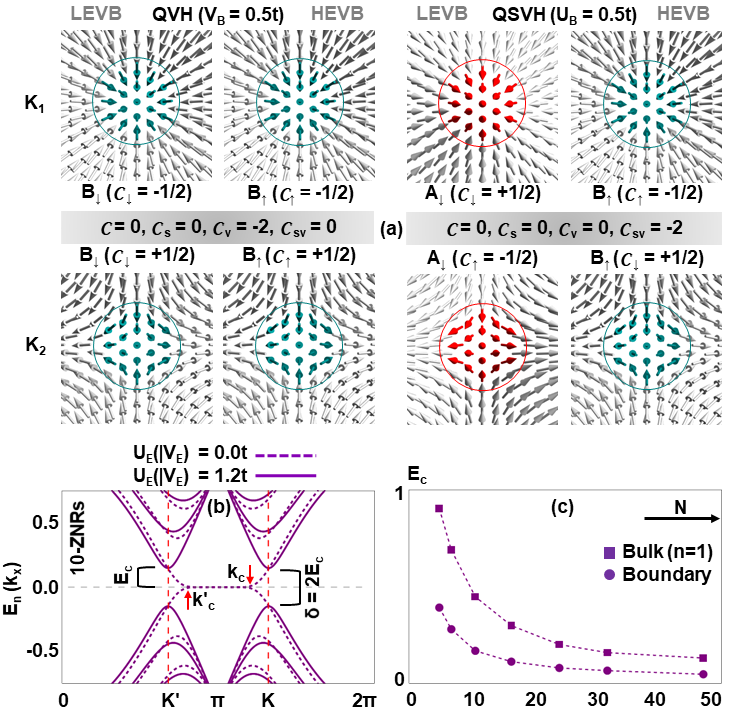}
\caption{\label{BBC-BD}\textbf{Bulk band topology and confinement effect.} \textbf{(a)} Meron/anti-Meron pseudospin textures of filled bands for QVH and QSVH phases. \textbf{(b)} Electronic spectrum of pristine ZNRs (dashed) and gapped ZNRs with boundary SPs (solid lines). Here $E_c$ is the energy of confined states on boundaries while $k_x=k_x^c$ and $k_x=k^{\prime c}_x$ are anti-crossing points in the edge state spectrum. \textbf{(c)} Confinement energy of boundary states and first bulk subband as a function of the ZNR width.}
\end{figure}

As shown in figure \ref{BBC-BD}(b), owning to the intrinsic band topology in the bulk \cite{kane-TI,SSH-Solitons,zak,AIII,Ryu02} and associated with the electronic wave functions localized on the boundaries \cite{GNR-Fujita,Nakada96, Katsunori99,Peres06}, nearest-neighbor hopping generates fourfold spin-degenerate energy-zero flat bands in the nontrivial regime of the 1D Brillouin zone connecting Dirac pints, $k\in({2\pi/3a_0,4\pi/3a_0})$. When the width of ZNRs is smaller than a critical limit, $W_z<W_z^c$, finite-size effects lead to confinement of both bulk and the boundary electronic states in the vicinity of Dirac points $k_x=K/K^{\prime}$. As shown in figure \ref{BBC-BD}(c), we plot the confinement energy of boundary states and the first bulk subband as a function of the ZNRs width. The energy of these confined states increases with decrease in the width. These tight binding results are consistent with the results obtained though low-energy effective Dirac theory for quantum confined ZNRs \cite{XNRs-Brey2,Muhammad-APR}.

The perturbations induced by SPs, both in the bulk and on the boundary, open an energy gap in the edge state spectrum. For instance, SSPs and SMPs in the bulk open an energy gap of $2V_B$ and $2U_B$, respectively, in the whole Brillouin zone. Such topologically nontrivial gap induces QVH and QSVH phase in the bulk and transforms gapless energy-zero flat bands into gapped flat bands on the boundaries. On the other hand, as depicted in figure \ref{BBC-BD}(b), SSPs and SMPs on the boundary have no effect on the bulk subbands but disperse energy-zero flat bands through momentum-dependent energy splitting in the edge state spectrum: energy of confined boundary states associated with momentum at valleys $k_x=K/K^{\prime}$ remain insensitive to SPs on the boundary but the energy of confined boundary states associated with time-reversal invariant momentum (TRIM) $k_x=\pi$ varies linearly with the strength of SPs on the boundary. Such momentum-dependent energy splitting leads to a confinement induced energy gap $\delta=2E_c$ at valleys. 

Though SPs both in the bulk and on the boundary open an energy gap in the edge state spectrum of quantum confined ZNRs, an interplay between SSPs in the bulk (on the boundary) and SMPs on the boundary (in the bulk) leads to SG-VF-CES, as shown in figure \ref{BBC}. For instance, in ZNRs with boundary magnetism ($U_E\ne0$), PEF-induced SSPs in the bulk evolves gapped edge state spectrum into SG-VF-CES. Similarly, in ZNRs with bulk magnetism ($U_B\ne0$), TEF-induced SSPs on the boundary evolves gapped edge state flat bands into SG-VF-CES.\par

Since spin remains a good quantum number and thus spin textures remain trivial in the absence of Rashba SOI, nontrivial topological character of SG-VF-CES in QVH and QSVH phases can be verified by examining the pseudospin textures of low energy bands. As shown in figure \ref{BBC-BD}(a), Meron and anti-Meron like pseudospin textures for filled/valence bands suggest that non-vanishing SSPs in the bulk lead to QVH state characterized by quartet $(\mathcal{C},\mathcal{C}_s,\mathcal{C}_v,\mathcal{C}_{sv})=(0,0,-2,0)$ while non-vanishing SMPs in the bulk lead to QSVH state characterized by quartet $(\mathcal{C},\mathcal{C}_s,\mathcal{C}_v,\mathcal{C}_{sv})=(0,0,0,-2)$. Here Chern, spin-Chern, valley Chern, and spin-valley Chern numbers are defined for a minimal four-band model as

\begin{equation}
\mathcal{C}=\mathcal{C}_{K_1}+\mathcal{C}_{K_2}=\left(\mathcal{C}_{K_1}^{\uparrow}+\mathcal{C}_{K_1}^{\downarrow}\right)+\left(\mathcal{C}_{K_2}^{\uparrow}+\mathcal{C}_{K_2}^{\downarrow}\right)\;.\label{C}
\end{equation}
\begin{equation}
2\mathcal{C}_s=C_{\uparrow}-C_{\downarrow}=\left(C_{\uparrow}^{K_1}+C_{\uparrow}^{K_2}\right)-\left(C_{\downarrow}^{K_1}+C_{\downarrow}^{K_2}\right)\;.\label{Cs}
\end{equation}
\begin{equation}
\mathcal{C}_v=\mathcal{C}_{K_1}-\mathcal{C}_{K_2}=\left(\mathcal{C}_{K_1}^{\uparrow}+\mathcal{C}_{K_1}^{\downarrow}\right)-\left(\mathcal{C}_{K_2}^{\uparrow}+\mathcal{C}_{K_2}^{\downarrow}\right)\;.\label{Cv}
\end{equation}
\begin{equation}
\mathcal{C}_{sv}=\mathcal{C}_{K_1}-\mathcal{C}_{K_2}=\left(\mathcal{C}_{K_1}^{\uparrow}-\mathcal{C}_{K_1}^{\downarrow}\right)-\left(\mathcal{C}_{K_2}^{\uparrow}-\mathcal{C}_{K_2}^{\downarrow}\right)\;.\label{Csv}
\end{equation}
where spin Chern number (\ref{Cs}) is subject to modulo 2. Since SPs on the boundaries do not affect the bulk band spectrum, but only negligibly small corrections in the vicinity of valleys, the band topology induced by SPs in the bulk remains intact even in the presence of SPs on the boundaries. Thus, similar to VF/SVF-CES in QVH/QSVH insulating phase, SG-VF-CES are characterized either by valley Chern number and valley-momentum locking or by spin-valley Chern number and spin-valley-momentum locking when the topologically nontrivial bulk energy gap is opened by SSPs or SMPs respectively. That is, if there is no inter-valley mixing, SG-VF-CES remain protected against backscattering. Further detail on bulk-boundary correspondence is discussed in supplementary information. \par

\begin{figure*}
\includegraphics[scale=0.65]{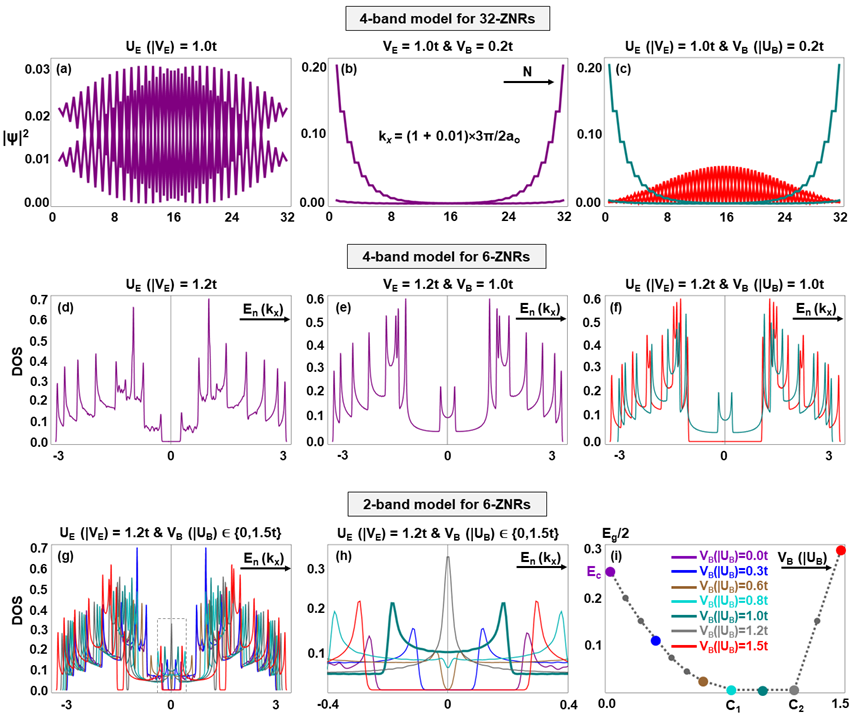}
\caption{\label{ES-DOS}\textbf{Edge state wave functions, density of states, and bulk-boundary gating effects.} \textbf{(a-c)} Edge state square wave functions in real space, at $k_x=K$, obtained from 4-band model for 32-ZNRs. SPs on the boundaries evolves exponentially decaying edge states into sinusoidal wave functions that correspond to gapped edge state spectrum (a). In the presence of SSPs in the bulk and that on the boundaries, exponentially decaying edge states wave functions imply gapless VF-CES (b). In the presence of SSPs (SMPs) in the bulk and SMPs (SSPs) on the boundaries, exponentially (sinusoidal) decaying edge states wave functions in spin down (up) sector imply SG-VF-CES (c). \textbf{(d-f)} Partial DOS obtained from 4-band model for 6-ZNRs. Vanishing DOS (d), non-vanishing DOS in both spin sectors (e), and non-vanishing (vanishing) DOS in spin down (up) sectors at energy-zero level confirms the distribution of edge state wave functions in real space. \textbf{(g-i)} Bulk-boundary gating effects on DOS obtained from low-energy 2-band model for 6-ZNRs. With fixed SPs on the boundaries ($U_E(|V_E)$), SPs in the bulk transform vanishing energy-zero DOS ($V_B(|U_B)<\xi_{c_1}$) into non-vanishing energy-zero DOS ($\xi_{c_1}<V_B(|U_B)<\xi_{c_2}$), and to vanishing energy-zero DOS ($V_B(|U_B)>\xi_{c_2}$). Here $\xi\in\{U,V\}$ and $\xi_{c_1}$ is the critical potential at which gapped edge states transform to gapless chiral edge state while $\xi_{c_2}$ is the critical potential at which gapless chiral edge states transform to gapped edge state. While $\xi_{c_1}$ depends upon the width of ZNRs, and thus, the confinement energy $E_c$, $\xi_{c_2}$ is equivalent to boundary SP, $\xi_{c_2}=U_E(|V_E)$. The symbol $|$ denotes `or' here.}
\end{figure*}

Consistent with the tight binding electronic dispersion, these results can be verified by studying real space distribution of edge state wave functions \ref{ES-DOS}(a-c) and DOS \ref{ES-DOS}(d-f). Sinusoidal behaviour of edge state wave functions \ref{ES-DOS}(a) and vanishing energy-zero DOS \ref{ES-DOS}(d), show that SPs on the boundary open an energy gap in the edge state spectrum. However, in the presence of SSPs in the bulk along with SSPs on the boundaries, edge state wave functions confined on boundaries \ref{ES-DOS}(b) and non-vanishing energy-zero DOS \ref{ES-DOS}(e) represent the existence of gapless VF-CES. Similarly, in the presence of SSPs (SMPs) in the bulk along with SMPs (SSPs) on the boundaries, exponentially (sinusoidal) decaying wave functions for spin-down (spin-up) edge states \ref{ES-DOS}(c) and no-vanishing (vanishing) energy-zero DOS for spin-down (spin-up) edge states \ref{ES-DOS}(f) confirm the existence of SG-VF-CES. That is, in the presence of SPs on the boundary, SPs in the bulk overcome confinement induced energy gap $\delta=2E_c$ in the edge state spectrum and ensure gapless CES are stemming from the nontrivial bulk-band topology.\par

As shown in figure \ref{ES-DOS}(i), unlike linear behaviour when the strength of bulk SPs exceeds the strength of SPs on the boundary, $\delta$ varies nonlinearly with the bulk SPs. It reminds that, similar to the nonlinear behaviour of boundary potentials \cite{Bhowmick10}, the dependence of $\delta$ on the bulk SPs also shows a Weber-Fechner type nonlinear behavior. However, there is a unique intermingling between confinement effects and the effects of SPs in the bulk and on the boundary. Since, $\delta$ remains independent of the strength of SPs on the boundaries, the minimum strength of bulk SPs desired for inducing SG-VF-CES is purely a function of ZNRs width. However, on the other hand, the minimum strength of boundary SPs required for the complete submergence of edge states with the bulk subbands at TRIM is independent of ZNRs width but remains dependent on the strength of bulk SPs. \par

The dependence of bulk SPs on the width of ZNRs and, in turn, dependence of boundary SPs on the bulk SPs can be understood as follows: Similar to the boundary states, energies of confined bulk states associated with momentum around TRIM remain independent of width. It implies that, in the absence of SSPs (SMPs) in the bulk, edge states at TRIM completely submerged with the bulk states when $U_E/t=1$ ($V_E/t=1$). However, in the presence of SSPs (SMPs) in bulk that induce a linear shift in the energies of bulk and boundary states across the whole Brillouin zone, the minimum strength of boundary SMPs (SSPs) required to submerge edge states with bulk subbands at TRIM increases to $U_E/t\ge1+2V_B/t$ ($V_E/t\ge1+2U_B/t$). On the other hand, since inter-edge hybridization between exponentially decaying edge state wave functions in the vicinity of valleys increases with decrease in the width, strength of bulk SPs required to overcome $\delta=2E_c$ also increases. In response, to meet the constraint on the boundary SPs to induce Dirac-like dispersion for SG-VF-CES at valleys and to ensure complete admixing of edge states with the bulk subbands at TRIM, the critical strength of boundary SPs also increases with decrease in the width. \par

The increase in critical strength of bulk SPs with decrease in width, and thus, the increase in critical strength of boundary SPs, demands a strategy to optimize this phenomenon in quantum confined ZNRs. Interestingly, as shown in figure \ref{BBC-1}, the critical strength of SMPs can be optimized though bulk-boundary gating schemes. This strategy also dictates how the topological half-metallicity characterized by SG-VF-CES is different from the PEF-driven conventional half-metallicity in the bulk spectrum \cite{Gong18} and a TEF-driven trivial half-metallicity in the edge state spectrum \cite{HMGNRs} of ZNRs. Furthermore, the generation of SG-VF-CES by a PEF (TEF) and the optimization of magnetic properties on the boundaries (in the bulk) through a TEF (PEF) opens a new direction for tuning magnetic properties via both PEF and TEF simultaneously. See further details in supplementary information.\par

\begin{figure*}
\includegraphics[scale=0.4]{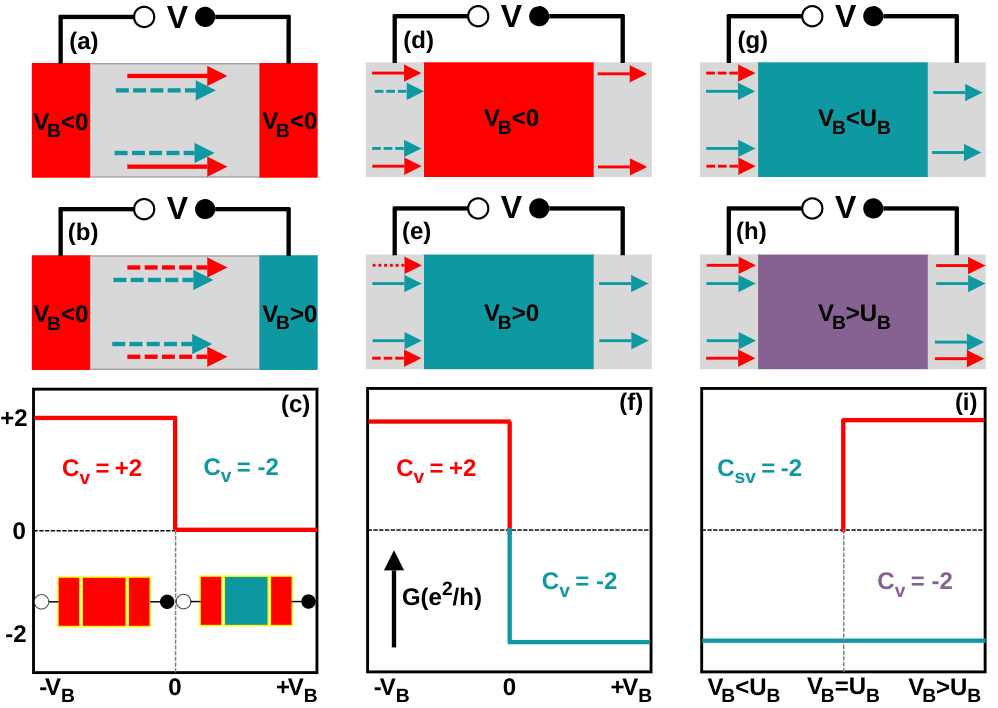}
\caption{\label{TopSpin}\textbf{Topological spintronic device applications} \textbf{(a-c)} Schematic representation for a spintronic device geometry where channel region in QVH insulating phase with VF-CES is contacted by topological Dirac spin-gapless phases in the on-state with same band topology of source and drain (a) and the off-state with different band topology of source and drain (b). Similar to spin-valve device configuration, quantized edge state spin conductance remain non-vanishing (vanishing) in the on (off) state (c). Topological switching of spin conductance is similar to the device configuration shown in the inset. \textbf{(d-f)} Schematic representation for a spintronic device geometry where channel region in topological Dirac spin-gapless phase is contacted by QVH insulating phase with VF-CES in the on-state (d) and off-state (e). Unlike topological switching at the drain of spin-valve geometry, topological switching in the channel region switches quantized edge state spin conductance from spin-up to spin-down in this configuration (f). \textbf{(g-i)} Schematic representation for a spintronic device geometry where channel region in topological Dirac spin-gapless phase is contacted by QVH insulating phase with VF-CES in the on-state (g) and off-state (h). Topological switching in the channel region switches quantized edge state spin conductance from SG-VF-CES to spin-degenerate VF-CES(i). Here solid (dashed) arrows represent gapless/conducting (gapped/non-conducting) chiral modes.}
\end{figure*}

Honeycomb lattice structures terminated on zigzag edges are special class of 2D materials where topological transport on 1D boundaries promise dissipationless topological electronics/spintronics applications \cite{phdthesis} with reduced subthreshold swing via Rashba effect \cite{Muhd-nano} and via negative capacitance \cite{NCTQFET} and may allow miniaturization owning to the intertwining of edge state transport with intrinsic band topology \cite{Muhammad-APR}. Such enhanced functionalities are associated with a strong intrinsic SOI in the ZNRs. Here we propose different mechanisation for topological spintronics, even in the absence of SOI. In a standard spintronic device geometry, figure \ref{TopSpin}, topological Dirac spin-gapless materials can be utilized both as spin injector/detector and spin-transport channel with large spin-relaxation lifetimes and spin-diffusion distances. \par

First, with a QVH phase hosting spin-degenerate VF-CES in the channel region, SG-VF-CES can be utilized as spin injector/detector. Second, with a QVH phase hosting spin-degenerate VF-CES at the source/drain contacts, SG-VF-CES can be utilized for topological edge state spin-transport in the channel region. Based on these two configurations, various spintronic logic devices such as spin-valve, magnetic tunneling junction, spin field effect transistor, and spin filter transistor can be modelled where giant magnetoresistance (GMR) is controlled by electric field driven topological switching of spin-polarization. As shown in \ref{TopSpin}, topological switching may be implemented either at the source/drain region or in the channel of a topological spintronic device geometry. In the former case, unlike utilizing magnetic field or spin-transfer torque, spin-current is controlled by electric field-driven topological switching of spin-polarization at the source or at the drain contact. In the later case, unlike Rashba SOI driven spin-precession, topological spin-transport via SG-VF-CES is enabled and manipulated through spin-valley locking with momentum, and that is implemented by a gate-controlled PEF in the channel.\par

In light of this, two different types of PEF-driven topological switching schemes can be realized in topological Dirac spin-gapless materials: (i) Topological switching between spin-down polarization characterized by $\mathcal{C}_v=-2$ and spin-up polarization characterized by $\mathcal{C}_v=+2$ in ZNRs with the boundary magnetism, as shown in figure \ref{TopSpin}(a-f). (ii) Topological switching between SG-VF-CES characterized by $\mathcal{C}_{sv}=-2$ to spin-degenerate VF-CES characterized by $\mathcal{C}_v=-2$ in ZNRs with the bulk magnetism, as shown in figure \ref{TopSpin}(g-i). See supplementary information for further discussion on the switching mechanisms, enhanced functionalities of proposed spintronic device configurations, and their comparison with previously proposed conventional/topological spin FETs. \par

In conclusion, we demonstrated the existence, practical realization, topological protection, and topological switching of SG-VF-CES that represent a novel topological Dirac spin-gapless phase. The proposed bulk-boundary gating scheme allows topological half-metallicity in ZNRs of 2D antiferromagnets, and that their magnetic properties may be practically tuned by a new strategy that exploits both PEF and TEF simultaneously. The proposed models may display several other advantages over previously proposed conventional/topological spin FETs and may guide a new strategy for engineering topological spintronic device applications where, unlike magnetic field, spin-transfer torque, or SOI driven switching, edge state topological spin-transport is enabled and manipulated via intrinsic spin-valley locking with momentum, and that is implemented by the gate-controlled PEF. Moreover, owning to a topologically protected Dirac dispersion, SG-VF-CES are well suited for steering the engineering of device applications with large Fermi velocity, high mobility, and large signal-to-noise ratio. Furthermore, the optimization of Weber-Fechner type nonlinear behavior in ZNRs may open a new direction for the miniaturisation of topological electronics/spintronics devices. The proposed mechanisms, studied for material-independent models here, are generic and applicable for a wide variety of strongly correlated and weakly spin-orbit coupled 2D materials, ranging from graphene to other 2D systems with honeycomb structure terminated on zigzag edges.\par

\begin{center}
\textbf{Methods}
\end{center}
Analytical and numerical tight binding calculations, based on symmetry analysis, are performed for zigzag nanoribbons of honeycomb lattice structure.

\begin{center}
\textbf{Data availability}\\
\end{center}
The datasets generated during and/or analysed during the current study are available from the corresponding author(s) on reasonable request.

\begin{center}
\textbf{Author contributions}\\
\end{center}
M.N. and X.W. contributed equally. M.N. performed the theoretical tight-binding calculations and wrote the manuscript with input from X.W.


\begin{center}
\textbf{Competing interests}\\
\end{center}
The authors declare no competing interests.

\begin{center}
\textbf{Additional information}\\
\end{center}
\textbf{Supporting material} includes further discussion on the bulk-boundary correspondence, optimization of magnetic properties via bulk-boundary gating scheme, and the PEF-driven topological switching.\\

\textbf{Correspondence and requests} should be addressed to M. Nadeem (mnadeem@uow.edu.au) or X. Wang (xiaolin@uow.edu.au).

\begin{acknowledgments}
This research is supported by the Australian Research Council (ARC) Centre of Excellence in Future Low-Energy Electronics Technologies (FLEET Project No. CE170100039) and funded by the Australian Government.
\end{acknowledgments}


\bibliography{apssamp}
\end{document}